\documentclass[12pt,preprint]{aastex}

\usepackage{graphicx}
\usepackage{tipa}

\slugcomment{The Astronomical Journal, Revised November 09}

\shorttitle{Active Asteroids}
\shortauthors{Jewitt}

\begin{document}

\title{Limits to  Ice on Asteroids (24) Themis and (65) Cybele}

\author{David Jewitt$^{1, 2}$  and Aurelie Guilbert-Lepoutre$^{1}$}
\affil{$1$ Dept. Earth and Space Sciences, UCLA \\
$2$ Dept. Physics and Astronomy, UCLA \\
}
\email{Contact: jewitt@ucla.edu}

\begin{abstract} 

We present optical spectra of (24) Themis and (65) Cybele, two large main-belt asteroids on which exposed water ice has recently been reported.  No emission lines,  expected from resonance fluorescence in gas sublimated from the ice, were detected. Derived limits to the production rates of water are $\lesssim$ 400 kg s$^{-1}$ (5$\sigma$), for each object, assuming a cometary H$_2$O/CN ratio.  We rule out models in which a large fraction of the surface is occupied by high albedo (``fresh'') water ice because the measured albedos of Themis and Cybele are low ($\sim$0.05 - 0.07).   We also rule out models in which a large fraction of the surface is occupied by low albedo (``dirty'') water ice because dirty ice would be warm, and would sublimate strongly enough for gaseous products to have been detected.  If ice exists on these bodies it must be relatively clean (albedo $\gtrsim$0.3) and confined to a fraction of the Earth-facing surface $\lesssim$10\%.  By analogy with impacted asteroid (596) Scheila, we propose an impact excavation scenario, in which 10 m scale projectiles have exposed buried ice.  If the ice is even more reflective (albedo $\gtrsim$ 0.6) then the timescale for sublimation of an optically thick layer can rival the $\sim$10$^3$ yr interval between impacts with bodies this size.  In this sense,  exposure by impact may be a quasi steady-state feature of ice-containing asteroids at 3 AU. 

 
\end{abstract}


\section{Introduction}
Exposed water ice on main-belt asteroids is thermodynamically unstable, on timescales that are very short compared to the age of the Solar system.  Ice is therefore not expected on the surfaces of asteroids, although it may be present underneath (Fanale and Salvail 1989).  Despite this, a growing number of observations suggest that ice does indeed exist at or very close to (within a thermal skin depth of) the physical surfaces of some main-belt asteroids.  For example, repetitive activity observed in two kilometer-scale, so-called ``main-belt comets'' at heliocentric distances $\sim$3 AU (133P/Elst-Pizarro and 238P/Read) is best explained by the sublimation of near surface ice (Hsieh et al.~2010, 2011).  However, no gas has been detected in these objects and they are too small (few km) and faint to allow the detection of water ice in near infrared reflection spectra (Licandro et al.~2011a, Rousselot et al. 2011).  Conversely, the 100-kilometer-scale asteroids (24) Themis and (65) Cybele show broad, near infra-red absorptions at $\sim$3.1 $\mu$m that are compatible in position and shape with vibrational bands in crystalline water ice (Campins et al.~2010, Rivkin et al.~2010, Licandro et al.~2011b).  These large, low (few percent) geometric albedo  asteroids raise questions about the stability and supply of ice in the asteroid belt, and about the possible connections between large, apparently ice-covered but inactive asteroids and small, mass-losing main-belt comets in which ice is suspected but undetected.

An alternative explanation for the 3.1 
$\mu$m band in terms of the mineral goethite (FeO(OH)) has been proposed by Beck et al.~(2011).  One difficulty with this interpretation is that, when found in meteorites, goethite is a product of aqueous alteration in the terrestrial environment. Extraterrestrial goethite in freshly-fallen meteorites is unknown (Alan Rubin, personal communication, 2011 July 01).   This could mean that goethite-rich asteroids exist but that there is no dynamical pathway for their collisionally produced fragments to reach the Earth as meteorites or, more simply, that there is no goethite in the asteroids.  (We note that the closely related mineral magnetite (Fe$_3$O$_4$) is found both in meteorites and in the spectra of asteroids (Yang and Jewitt 2010). It is not obvious how magnetite-containing meteorites could find their way to Earth while goethite-containing ones could not).  Regardless, the interpretation of the 3.1 $\mu$m band in Themis and Cybele as due to surface ice remains plausible.

The bulk density of C-type asteroids with diameters in the 100 km - 300 km range is $\rho$ = 2000 kg m$^{-3}$, albeit with a large uncertainty (Baer and Chesley 2008).  With this density, the gravitational escape speeds from Themis and Cybele are $V_e$ =  105 m s$^{-1}$ and 140 m s$^{-1}$, respectively.  In both cases, $V_e$ is small compared to the $\sim$750 m s$^{-1}$ thermal speeds of water molecules at temperatures ($\sim$180 K) appropriate to low albedo asteroids in the outer  belt, so that sublimated molecules should flow unimpeded into space.  This leads us to ask whether evidence might be found for gas produced by sublimation of the ice from these bodies.    
In this short paper, we describe spectral line observations taken in search of this gas.

\section{Observations}
The strongest limits to the production of gas in comets are placed using optical spectra of daughter molecules (fragments of stable parent molecules produced by photodissociation).  The brightest daughter emission line in the optical spectra of comets is the OH 3080\AA~band.  Unfortunately,  this band lies near the ozone atmospheric absorption edge making ground-based observations of OH consequently difficult, while no suitably sensitive space-based instruments exist.  Spectroscopic detections of gas in comets are more usually made using the second-brightest line, produced by resonance fluorescence of the CN radical and located near 3889\AA~(Schleicher 2010).  Whereas OH is produced as a photodissociation product of H$_2$O, the source of CN is less certain.  It is at least in part a photodissociation product of HCN, but may also be released from solid grains (e.g. see recent discussion in Paganini et al.~2010).

Accordingly, we took spectroscopic observations targeting outgassed CN from Themis and Cybele, on UT 2010 September 11 at the Keck I 10-meter  telescope atop Mauna Kea, Hawaii.  We used the Low Resolution Imaging Spectrometer (LRIS, Oke et al.~1995), which is equipped with two 4096$\times$4096 pixel charge-coupled device detectors producing an image scale 0.135$\arcsec$ per pixel and fed by a dichroic filter.  We employed the ``460''  dichroic, which has 50\% transmission at 4900\AA, and a 1$\arcsec$ wide slit for all observations.   The geometrical circumstances of the observations are given in Table (\ref{geometry}) while the orbital properties of Cybele and Themis are listed in Table (\ref{orbits}).  

Themis and Cybele were identified using finder charts and confirmed by their motion relative to the fixed stars.  Integrations of 400 s and 330 s were secured on Themis and Cybele, respectively, together with spectra of nearby solar analogs HD209847 and Hyades 64.   During all observations we used an image rotator  to hold the long axis of the 1.0$\arcsec$ wide slit perpendicular to the local horizon, thereby minimizing effects due to atmospheric dispersion.  We guided the telescope by hand using reflection from the spectrograph slit jaws to keep the target objects correctly positioned within the slit.  For each new pointing of the telescope, we secured wavelength and flat-field calibration spectra using a set of lamps internal to LRIS.  

\begin{figure}
\epsscale{0.9}
\begin{center}
\plotone{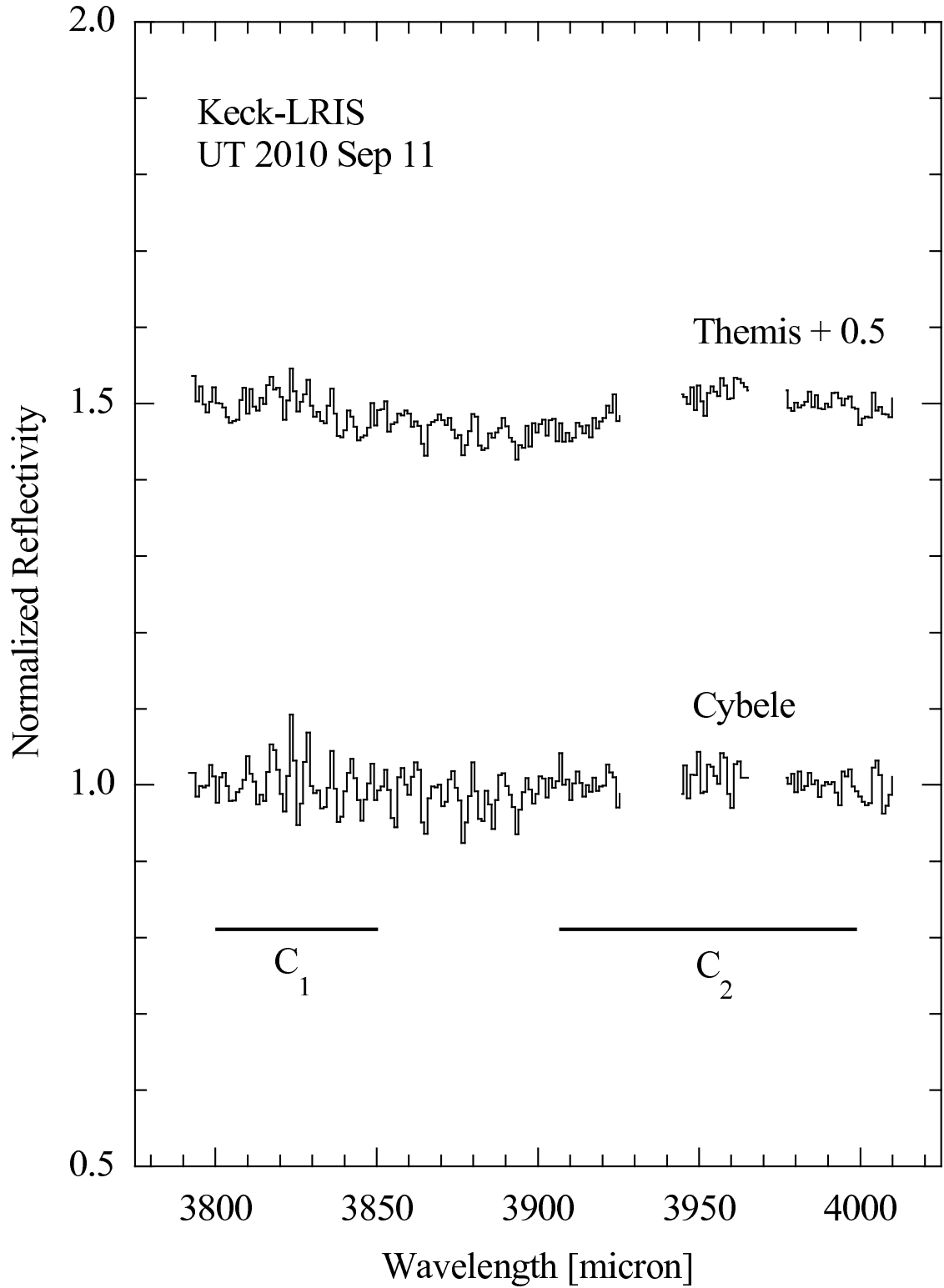}
\caption{Normalized reflectivity spectra of Themis  and Cybele showing the two wavelength regions, $C_1$ and $C_2$, used to define the CN continuum.  The spectrum of Themis has been displaced vertically by 0.5 for clarity.    \label{spectra} } 
\end{center} 
\end{figure}

Data reduction was performed by flattening the wavelength-calibrated spectra and extracting the signal from a region along the slit 6.75$\arcsec$ in length.  Reflectivity spectra were computed by dividing each object spectrum by the spectrum of a solar analogue and normalizing the result to unity at 3900\AA.  The normalized reflectivity spectra are shown in Figure (\ref{spectra}), where they are  offset vertically for clarity of presentation.  Themis and Cybele are spectrally featureless at the wavelengths of CN.  

We estimate a limit to the gas production rate using the procedure described in Jewitt et al.~(2009).  An upper limit to the CN flux density, $f_{CN}$, is obtained from 

\begin{equation}
f_{CN} = 5 \sigma_c \left[\frac{S_{3880}}{S_B}\right] \left[\frac{f_{3880}}{f_B}\right]_{\odot} f_{B} W_{CN}.
\label{fcn}
\end{equation}

\noindent Here, $\sigma_c$ is the (dimensionless) uncertainty on the normalized continuum at the wavelength of the CN band, $f_B$ and $f_{3880}$ (erg cm$^{-2}$ s$^{-1}$ \AA$^{-1}$) are the  flux densities at the wavelengths of the  B-filter and at 3880\AA, respectively, and $W_{CN}$ =40\AA~is the width of the portion of the spectrum in which the CN emission is sought.  Quantity $S_{3880}/S_{B}$ is the ratio of the reflectivities at the wavelengths of CN and the B-filter.  Both asteroids are nearly spectrally neutral (Table \ref{physical}) and we set $S_{3880}/S_{B}$  = 1.  Quantity $(f_{3880}/f_{B})_{\odot}$ = 0.52 is the ratio of the flux densities at the two specified wavelengths in the Solar spectrum (Arvesen et al.~1969).  The factor of five corresponds to our deliberately conservative use of a  5$\sigma$ confidence limit.   

We first computed $\sigma_c$, the standard deviation on the reflectivity in two continuum windows flanking the CN band.    We used the continuum windows $C_1$ = 3800 - 3870\AA~and $C_2$ = 3910 - 4000\AA~and omitted the wavelengths of the solar H and K lines of Calcium since these were imperfectly cancelled by the computation of the reflectivity spectrum.  To measure $\sigma_c$ we first  removed residual trends by fitting a 2nd order polynomial to the 3800~\AA~$\le \lambda \le$~4000~\AA~region, excluding wavelengths from 3870 - 3910\AA~and divided the reflectivity by the polynomial.   For Themis and Cybele, we obtained $\sigma_c$ = 0.0026 and $\sigma_c$ = 0.0027, respectively.  

We determined $f_{B}$ by a self-calibration method, as follows.  The 1$\arcsec$ slit excludes a significant but unknown fraction of the light from the asteroid, and so the signal in the spectrum cannot be used directly to secure accurate photometry with reference to the standard stars.  However, the fraction of the total light excluded by the slit is approximately independent of wavelength, since the point-spread-function of LRIS is constant and since the use of an image rotator precludes the possibility of differential refraction out of the slit.  Given this, we use the flux density calculated from the broadband magnitudes (Table \ref{physical}) to estimate the flux density at the wavelength of the CN band. This method is subject to errors in the published B-filter magnitudes owing to, for example, possible rotational lightcurves of the objects.  However, both objects have small rotational variations ($\sim$0.09 magnitudes with a rotation period of 8.3744$\pm$0.0002 hr for Themis (Harris et al.~1989) and $\sim$0.06 magnitudes and 6.07$\pm$0.05 hr for Cybele (Schober et al.~1980)) and in neither case is rotation a likely cause of significant photometric error.  As a result, we feel that the uncertainties in $f_{B}$ in this self-calibration method are probably no more than a few tens of percent, and certainly smaller than a factor of two, which is sufficient for our present purposes.  The final 5$\sigma$ limits to the fluxes from Themis and Cybele are $f_{CN}$ = 7.5$\times$10$^{-15}$ erg cm$^{-2}$ s$^{-1}$ and 6.1$\times$10$^{-15}$ erg cm$^{-2}$ s$^{-1}$, respectively.


The CN band is produced by resonance fluorescence of sunlight, in which cometary molecules scatter photons randomly in direction but without change in energy.  When the coma is optically thin, the scattered flux, $f_{CN}$ [erg cm$^{-2}$ s$^{-1}$], received at distance $\Delta$ [cm] is simply proportional to the number of CN molecules, $N$, in the section of the slit from which the spectrum was extracted, or

\begin{equation}
f_{CN} = \frac{g(R) N}{4 \pi \Delta^2}.
\label{fluorescence}
\end{equation}

\noindent The constant of proportionality in Equation (\ref{fluorescence}) is called the resonance fluorescence efficiency, $g(R)$ [erg s$^{-1}$ molecule$^{-1}$].  It is conventionally evaluated at heliocentric distance $R$ = 1 AU and scaled to other distances using $g(R)$ = $g(1)$/$R^2$.  At a given $R$, the fluorescence efficiency of CN can vary by a factor of $\sim$1.5, owing to the Doppler shift between cometary lines and the Solar spectrum (the so-called Swings effect).   On UT 2010 Sep 11, Themis and Cybele had radial velocities $dR$/$dt$ = +0.7 km s$^{-1}$ and +1.1 km s$^{-1}$, for which the fluorescence efficiencies are, respectively, $g(1)$ = 2.8$\times$10$^{-13}$ [erg s$^{-1}$ molecule$^{-1}$] and $g(1)$ = 2.9$\times$10$^{-13}$ [erg s$^{-1}$ molecule$^{-1}$] (Schleicher 2010). 

Substituting into Equations (\ref{fcn}) and (\ref{fluorescence}), with $R$ and $\Delta$ drawn from Table \ref{orbits} for UT 2010 Sep 11, we obtain $N$ = 7.7$\times$10$^{27}$ CN molecules in the projected slit at Themis and $N$ = 8.1$\times$10$^{27}$ CN molecules at Cybele.  To estimate the production rate from emission line data we used the Haser model, in which the parent of CN decays with a length scale (at $R$ = 1 AU) $\ell_P$ = 1.3$\times$10$^4$ km and the CN daughter itself is photo-destroyed on the length scale $\ell_{d}$ = 2.1$\times$10$^5$ km (A'Hearn et al.\ 1995).  We integrated the Haser number density model over the projected spectrograph slit and assumed that the outflow speed of the sublimated gas is $v$ = 0.75 km s$^{-1}$, consistent with measurements in comet C/Hale-Bopp at about this heliocentric distance (Biver et al.\ 2002). For Themis, the resulting limit to the production rate of CN is $Q_{CN} <$ 3.6$\times$10$^{25}$ s$^{-1}$. 
For Cybele, the corresponding limit is $Q_{CN} <$ 3.5$\times$10$^{25}$ s$^{-1}$.  

We are interested in the mass-dominant volatile H$_2$O, not CN.  In other comets, the average ratios of the production rates of CN to OH and of OH to H$_2$O are $Q_{OH}$/$Q_{CN} \sim$ 320 and $Q_{OH}$/$Q_{H_2O}$ = 0.9, respectively, giving $Q_{H_2O}$/$Q_{CN} \sim$ 360 (A'Hearn et al.\ 1995).  If this ratio applies to Themis and Cybele, the implied limits to the mass production rates in water are $Q_{H_2O} <$ 1.3$\times$10$^{28}$ s$^{-1}$, corresponding to $<$ 390 kg s$^{-1}$ for Themis and $Q_{H_2O} <$ 1.2$\times$10$^{28}$ s$^{-1}$, corresponding to $<$ 370 kg s$^{-1}$ for Cybele.  The accuracy of these limits is subject to the assumption that $Q_{H_2O}$/$Q_{CN}$ has a value similar to that measured in comets.  We note, however, that Lovell et al.~(2010) independently used UV and radio-line observations to directly obtain an upper limit $Q_{OH} <$ 1$\times$10$^{28}$ s$^{-1}$ (3$\sigma$) at Themis, corresponding to 300~kg~s$^{-1}$.  This value is consistent with the limits derived here based on CN.  





\section{Discussion}

\subsection{Simple Model}


We seek to estimate the fraction of the surface on each asteroid that could be occupied by exposed water ice, reasoning as follows.  The upper limit to the mass loss rate from our spectra sets an upper limit to the area of exposed ice on the surface.  The limit is a function of the unknown albedo of the ice, with dirty (low albedo, warm) ice confined to a smaller possible area than clean (high albedo, cold) ice.  Separately, the mean albedos of the asteroids set an independent albedo-dependent constraint on the area of exposed ice, since a large area of high-albedo ice, even when averaged with the lower albedo of surrounding non-ice surface, would violate the low measured albedos.  The two observational constraints, on mass loss and on the average albedo, limit both the possible ice albedo and the fraction of the surface covered by ice.  



If a fraction of the surface, $f$,   is covered in ice of albedo $p_i$ and the remaining fraction, $1-f$, is covered in non-ice (or ``dirt'') of albedo $p_d$, we may write the average albedo as $\overline{p} =  \left[ f p_i + (1-f) p_d \right]$. Solving for $f$ gives

\begin{equation}
f = \frac{\overline{p} - p_d}{p_i - p_d}
\label{f}
\end{equation}

%

\begin{figure}
\epsscale{0.9}
\begin{center}
\plotone{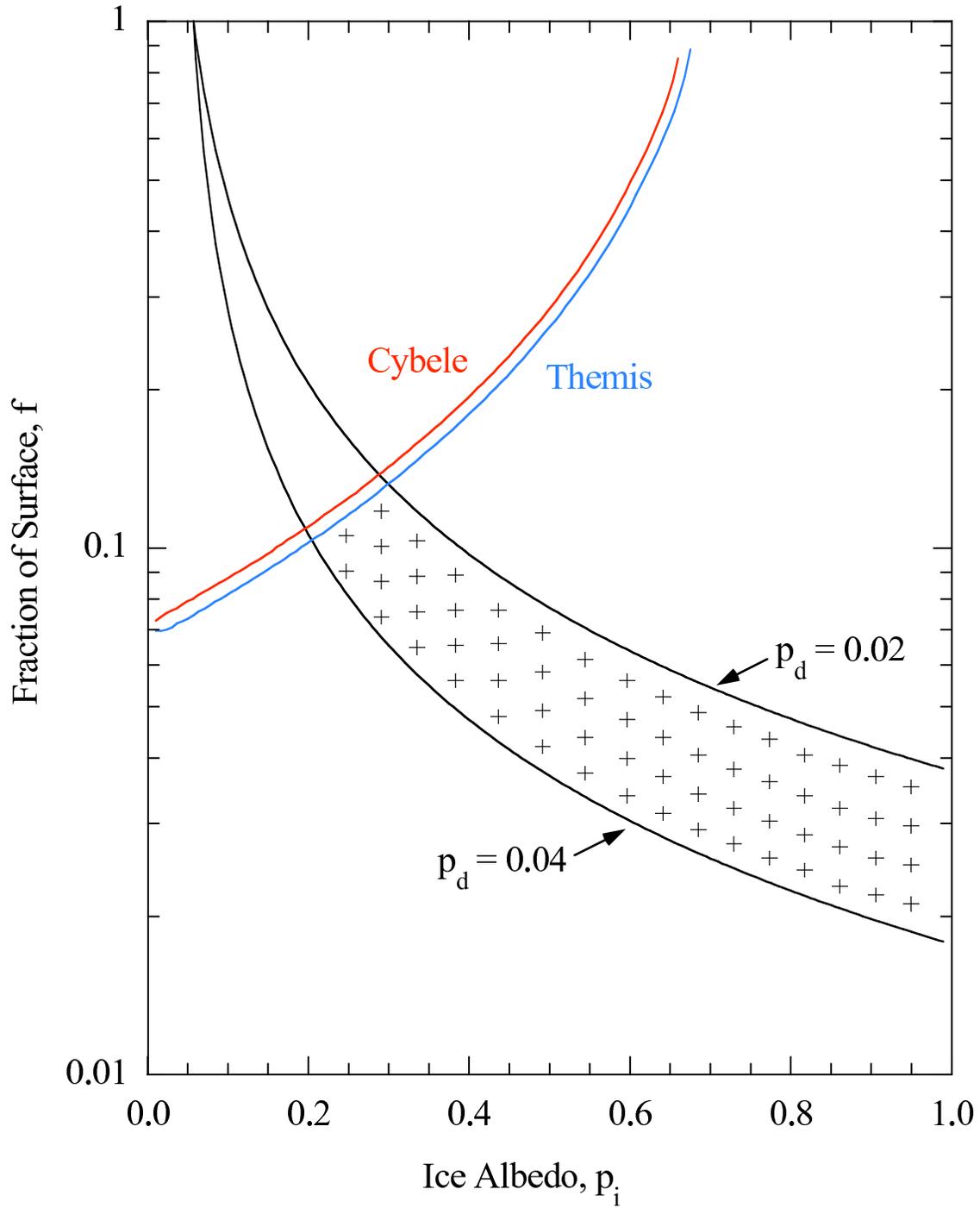}
\caption{Allowable values of the ice albedo are plotted against the fraction of the surface occupied by ice.  Black lines reflect the albedo constraint (Equation \ref{f}).  Red and blue lines show the sublimation constraint for Cybele and Themis, respectively.  The shaded region shows allowable solutions for the two asteroids.   \label{model} } 
\end{center} 
\end{figure}

\noindent Two black lines in Figure (\ref{model}) show Equation (\ref{f}) with two assumed values of the non-ice albedo, $p_d$ = 0.02 and 0.04, to represent the least reflective asteroid surfaces known.  Acceptable solutions must lie along or between the trajectories defined by these curves in order to guarantee that the area-averaged albedo equals the measured value (from Equation (\ref{f})).  

\begin{figure}
\epsscale{0.9}
\begin{center}
\plotone{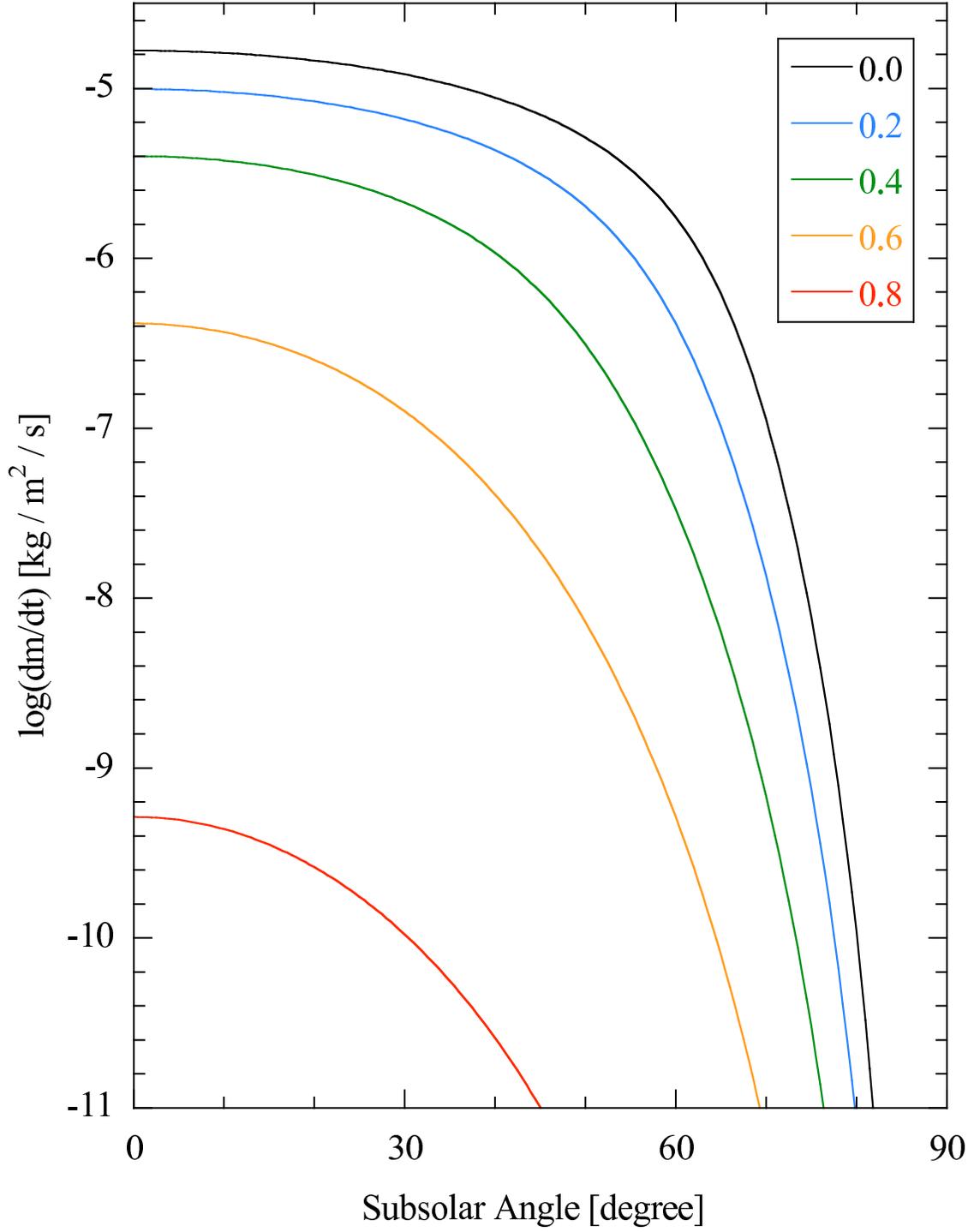}
\caption{Thermal equilibrium sublimation mass flux versus angle from the sub-solar point, computed for five different values of the ice albedo, as marked.  \label{dmbdt} } 
\end{center} 
\end{figure}

To compute a second constraint from the absence of detected gas, we solved the heat balance equation on the surface of a spherical model asteroid having the instantaneous location of Themis and Cybele as in Table (\ref{geometry}).  The equilibrium sublimation rate is given by solution of

\begin{equation}
\frac{L_{\odot}}{4 \pi R^2}(1-A) \cos(\theta) = \epsilon \sigma T^4 + L(T) \frac{dm}{dt}(\theta, T)
\label{sublimation}
\end{equation}

\noindent in which $L_{\odot}$ is the luminosity of the Sun, $R$ is the heliocentric distance in meters, $A$ is the Bond albedo, $\theta$ is the angle between the direction to the Sun and the normal to the surface, $\epsilon$ is the emissivity of the body, $\sigma $ is the Stefan-Boltzmann constant and $L(T)$ is the latent heat of sublimation of water ice at temperature, $T$.  The quantity $dm/dt$ is the sought-after mass flux of sublimated ice.   The two terms on the right represent power radiated from the surface into space and power used to break hydrogen bonds in the sublimation of ice.  The term on the left represents power absorbed from the Sun.  Equation (\ref{sublimation}) ignores a term to account for conduction of heat into the interior, but this neglect is justified (in the case of Cybele, at least) by detailed thermophysical models in which the thermal inertia is found to be a very small  $I$ = 15  J m$^{-2}$ s$^{-0.5}$ K$^{-1}$ (Mueller and Blommaert 2004). This low $I$ corresponds to a highly porous surface material in which heat conduction will be negligible.   Equation (\ref{sublimation}) cannot be solved explicitly, since it contains two unknowns, namely $T$ and $dm/dt$.  We used experimental determinations of the temperature dependence of $L(T)$ and of $dm/dt$ from Washburn (1926) and solved Equation (\ref{sublimation}) numerically for each element of the surface on a sphere located at the distance of each asteroid.  Figure (\ref{dmbdt}) shows sample solutions for $dm/dt$ as a function of angular distance from the subsolar point, on a model nucleus in which the spin vector points at the Sun.  Again, in the case of Cybele, a sun-pointing spin vector is justified by a detailed analysis.  The geocentric ecliptic latitude and longitude of Cybele on 2010 Sep 11 were $\lambda$ = 53$\degr$, $\beta$=-4$\degr$, respectively, which lies within 21$\pm$15$\degr$ of the pole solution reported by Mueller and Blommaert 2004.  The spin vector of Themis is unknown, but time-resolved spectra over one night show no temporal variation (Campins et al.~2011), consistent with a nearly pole-on perspective.

Figure (\ref{dmbdt}) shows that the subsolar, thermal equilibrium sublimation flux for $A$ = 0 is about 10$^{-5}$ kg m$^{-2}$ s$^{-1}$, but this falls by $\sim$4.5 orders of magnitude as the albedo rises to $A$ = 0.8 and also drops precipitously at $\theta >$80$\degr$.  The stability of the ice evidently depends very strongly on whether it is ``clean'' (high albedo) or ``dirty'' (low albedo) and on the position of the ice relative to the subsolar point (Figure \ref{dmbdt}).  The difference between dirty and clean ice can, in this context, be the result of the admixture of very small mass fractions of absorbing material. For example, $\ll$ 1\% contamination by mass can, depending on the grain size and composition of the contaminant, depress the optical albedo of ice to $\le$0.1 (Clark 1982).  

We computed the maximum possible ice-covered surface fraction by integrating $dm/dt$ over the coldest areas on the model asteroid, from $f = (1- \sin(\theta_c))$, with $\theta_c$ defined by

\begin{equation}
\frac{dM}{dt} = \int_{\theta_c}^{\pi/2}2 \pi r_n^2  \sin(\theta) \frac{dm}{dt}(\theta,T) d\theta.
\label{total}
\end{equation}

\noindent In Equation (\ref{total}), $dM/dt$ (kg s$^{-1}$) is the total mass production rate (which we set equal to the upper limit values obtained from the data) and $\theta_c$ is the critical subsolar angle, such that sublimation integrated over all regions with $\theta \ge \theta_c$ equals $dM/dt$.  The asteroid radius, $r_n$, is taken from Table (\ref{physical}).

Figure (\ref{model}) shows the sublimation constraints graphically in terms of the fraction of the surface occupied by ice as a function of the albedo of the ice.  Red and blue curves show the constraints imposed by the non-detection of sublimated molecules for Themis and Cybele, as determined by Equations (\ref{sublimation}) and (\ref{total}).  Acceptable solutions must lie on or below the plotted red and blue curves, otherwise gas production by sublimation would have been large enough to be detected.  The region satisfying both constraints is shown shaded.  Figure (\ref{model}) shows that models in which  ice is globally distributed ($f$ = 1) on Themis and Cybele are excluded.  At one extreme, this is because full-surface clean ice would violate the (low) average albedos. At the other, full-surface dirty ice would be warm enough to strongly sublimate and would violate the spectroscopic constraint on sublimated gas.  No intermediate albedo solution with $f \sim$ 1 is found.  Instead, the allowed solutions for both asteroids are those with ice albedos $p_i \gtrsim$ 0.3 and coverage fractions $f \lesssim$ 0.1.  Ice on these bodies, if it is present, must be relatively clean and spatially localized.  

Measurements from a single rotation show that the ice band on Themis is approximately constant in depth (Campins et al.~2010), consistent with a small angle between the line of sight and the spin vector.  Independent observations dispersed over a five year interval show a larger variation (e.g.~using spectra reported in the Supplement to Rivkin and Emery~2010 we measured the fractional band depth in 2005 to be $\sim$40\% deeper than in 2003 (continuum linearly interpolated between 2.7 and 3.5 $\mu$m).  Some of this variation is no doubt instrumental, but the data are consistent with a substantial variation).  In the future, a determination of the spin vector of Themis will be of great value in further localizing the ice on this body.  Spin vector solutions have been published for Cybele (Muller and Blommaert 2004) but  3 $\mu$m spectra have been reported from only two nights (UT 2009 Sep. 9; Licandro et al.~2011), insufficient to determine the spatial distribution.

It is interesting to note that the 3.1 $\mu$m band appears in Themis and Cybele at only $\sim$10\% of the local continuum depth.  This band is intrinsically very strong (absorption coefficient $k \sim$ 10$^{6}$ m$^{-1}$; Irvine and Pollack 1968) and frequently appears saturated in astronomical spectra.  The subdued band depth is consistent with a 10:1 dilution of the ice spectrum with bland (dust) continuum and so matches the inference drawn above, that the ice surface coverage fraction is $f \lesssim$ 0.1. Rivkin and Emery (2010) took a different approach, arguing that the unsaturated band results from a very limited optical path length through the ice.  They used ice coatings only $\sim$45 nm in thickness to model the spectrum of Themis, apparently assuming $f$ = 1, and were able to match the band depth and shape in detail.  The authors speculated that a thin, global ice film could result from freezing of water vapor emanating from beneath the surface.  

However, the thin, global ice film hypothesis has two  problems.  First, the radial temperature gradients on the  sun-facing hemispheres of Themis and Cybele are in the wrong direction for surface frost to form.  The dayside surfaces are warmer than the subsurface regions so that any free water molecules would migrate towards, and freeze at, deeper layers, not at the hot surface.  Second, even if a 45 nm ice layer could form (by freeze out on the night side, for example), its lifetime to sublimation would be very short, requiring an unreasonably large water mass flux from each asteroid in order to maintain steady state.  To see this, we first note that a 45 nm ice coating is too thin to have a substantial effect on the albedo, as shown by both models and measurements of non-absorbing coatings on grains (e.g. Lasue et al.~2007). As a result, the temperature of the ice film would be set by the temperature of the grain material with which it sits in close physical contact.  For low Themis-like albedos, Figure (\ref{dmbdt}) shows that the subsolar sublimation rate is of order $dm/dt$ = 10$^{-5}$ kg m$^{-2}$ s$^{-1}$, falling to $dm/dt$ = 10$^{-6}$ kg m$^{-2}$ s$^{-1}$ at a subsolar angle of 60$\degr$.   At these rates, an ice film 45 nm in thickness would sublimate away on timescales of only 5 s to 50 s over most of the sunlit surface of the asteroid.  Rivkin et al.~(2010) and Campins et al.~(2010) assert that the ice is widespread or global, meaning that the total mass loss rate due to sublimation from an $r_n$ = 100 km radius Themis or Cybele-like asteroid should be $4 \pi r_n^2 dm/dt$ = 10$^5$ to 10$^6$ kg s$^{-1}$, in steady state.  These rates are two to three orders of magnitude larger than our spectroscopic limits to the gas production  ($<$390 kg s$^{-1}$ and $<$370 kg s$^{-1}$ for Themis and Cybele, respectively) and so are observationally ruled out.  Therefore, we conclude that the thin, global ice film hypothesis is inconsistent with the data.  Another explanation is needed.

\subsection{Excavation Model}


Themis and Cybele are too small to have retained primordial heat, and too small to be significantly heated by the decay of long-lived radioactive elements trapped within their constituent rocks.  Unlike some planetary satellites (for example, Enceladus) they also lack any source of flexural heating driven by external periodic gravitational forces.  Therefore, we reject the possibility that surface ice could represent  water vapor frozen on the surface after being driven out from a warm interior by internally-driven fumarolic activity.


Recent observations of another large asteroid suggest a different possibility.   The 113 km diameter (596) Scheila displayed transient activity after being struck by a body several tens of meters in diameter (Jewitt et al.~2011, Ishiguro et al.~2011 while Bodewits et al.~2011 deduced a somewhat larger impactor).  Dust ejected from Scheila was observed to disperse under the action of solar radiation pressure on timescales of weeks, while the near infrared ice bands were not detected  (Yang and Hsieh 2011). We suggest that Themis and Cybele are analogs of (596) Scheila in that they were also struck by small projectiles but differ from Scheila in that the escaping ejecta has long-since departed and that the impacts excavated ice.  In all three cases, impact ejecta was then deposited on the surface as fallback debris, perhaps creating ray-like systems on each body.  Sub-surface ice can survive against sublimation at $\sim$3 AU  when insulated by even meter-thick layers of refractory regolith (Schorghofer 2008).  As we will discuss momentarily, craters of the size envisioned for Themis and Cybele would easily puncture meter-thick refractory surface layers.  It is thus reasonable to conjecture that some impacts might excavate buried ice in this way.

We first need to estimate the size of the projectile needed to expose ice sufficient to cover $\sim$10\% of the surface.  We assume a reference thickness of the impact deposited ice as $\ell \sim$ 1 mm, since an ice or frost layer much thinner than this will not strongly affect the albedo.  Then, the volume of the exposed ice is $V \sim f \pi r_n^2 \ell$, where $r_n$ is the asteroid radius.  Most of the ejecta from impact at 5 km s$^{-1}$ onto a 100 km scale body is launched below the escape speed (Housen and Holsapple 2011).  We therefore  assume that the ice volume is comparable to the crater volume, $V_c = 2 \pi r_c^3/5$, where we approximate the crater by  a cylinder with radius $r_c$ and depth to diameter ratio 1/5.  Setting $V = V_c$, we find

\begin{equation}
r_c = \left(\frac{5}{2}f r_n^2 \ell \right)^{1/3} 
\label{crater}
\end{equation} 

\noindent for the radius of the impact crater needed to produce enough ejecta to cover a fraction, $f$, of the surface with ice to a depth, $\ell$.  Substituting $f$ = 0.1, $r_n \sim$ 100 km and expressing $\ell$ in mm, we find $r_c \sim$ 130 $[\ell/(1 mm)]^{1/3}$ m.  For $\ell$ = 1 mm, such a crater would occupy $\lesssim$10$^{-6}$ of the surface area of a 100 km radius target body.


In the gravity-scaling limit,  the radius of the projectile needed to form a crater with $r_c$ = 130 m on a 100 km radius asteroid at impact speed 5 km s$^{-1}$ is $\sim$1 m, assuming normal impact and identical target and projectile densities.  However, experiments with impact into porous, fragmental and icy materials show that a larger projectile is needed to form a crater of a given size, relative to the gravity-scaled estimate (e.g. Holsapple 1993, Arakawa et al.~2000).  For example, Equation (1) of the latter paper gives a projectile radius $\sim$5 m for normal impacts into porous, icy material, with other parameters held the same.  The surface conditions on Themis and Cybele are unknown but it is quite likely that these bodies possess a fragmental surface layer caused by past impacts and statistically likely that the impacts were not perpendicular to the local surface, so that the projectiles were larger than indicated by the gravity-scaling estimate.  

To proceed, we consider a nominal impactor radius of 10 m, on the understanding that this size is probably slightly too large but is in any case uncertain by a factor of several as a result of the unknown impactor speed, density, incident angle and impact parameter.  We used the asteroid collision probabilities from Bottke et al.~(1994) in order to estimate $\tau_c$, the interval between impacts of 10 m projectiles onto a 100 km radius target asteroid.  These collision probabilities are based on measurements of asteroids larger than 1 km in size, requiring a factor of 100 extrapolation to reach the 10 m  scale of the projectiles implicated here.  Unfortunately, knowledge of the size distribution of sub-kilometer asteroids in the main-belt is limited because most such bodies are too faint to be directly observed.  We rely on crater counts from asteroid Gaspra to provide an indication of the size distribution of small projectiles.  There, impact craters from 0.4 km to 1.5 km in diameter are distributed as a power-law with a differential size index -3.7$\pm$0.5 (Belton et al.~1992, while fresh craters may follow a slightly steeper index according to Chapman et al.~1996).  Craters in this size range result from the impact of projectiles a few decameters in radius, again subject to significant uncertainties about the surface physical properties of Gaspra.  Using this size distribution, we estimate that the timescale for impact onto Themis or Cybele-sized asteroids is $\tau_c \sim$10$^3$ yr.  This timescale is clearly very uncertain because of the large extrapolation from the relatively well-sampled asteroids at kilometer size-scales down to the 10 m sizes of the projectiles.   For example, the estimated numbers of asteroids with $r_n \ge$ 5 m range from $\sim$10$^{10}$ to $\sim$10$^{12}$ (Davis et al.~2002), indicating a two order-of-magnitude  uncertainty in $\tau_c$ at small projectile sizes.  With these caveats in mind, we consider the implications of a 10$^3$ yr timescale for impact.

First, we note that the number of non-overlapping craters that can be placed on the surface is roughly $N_c \sim 4 r_n^2/(r_c^2)$.  With $r_n$ = 100 km and $r_c$ = 130 m, we find $N_c \sim$ 2$\times$10$^6$, corresponding to the accumulation of impacts over $\sim$2 Gyr.  Buried ice could persist against repeated impact excavation over a large fraction of the age of the solar system.

We next determine the conditions which must prevail for excavated ice to survive on the surface for $\tau_c \sim$ 10$^3$ yr.  The mass column density of a 1 mm thick ice layer is $\rho \ell$ = 1 kg m$^{-2}$.  If this layer is to survive for 10$^3$ yr  (3$\times$10$^{10}$ s), the mean loss rate cannot exceed $\rho \ell/\tau_c \sim$ 3$\times$10$^{-11}$ kg m$^{-2}$ s$^{-1}$, setting an upper limit to the equilibrium sublimation temperature $T_c \le$ 131 K from Equation (\ref{sublimation}).  In the specific model considered here, with the Sun on the projected rotation axis, sufficiently low temperatures are found in a thin band around the equator where $\cos(\theta)$ in Equation (\ref{sublimation}) is suitably small.  By analogy with the Moon, surface ice on Themis and Cybele might survive best in regions protected from Sunlight by local topography.

\subsection{Numerical Thermal Model}

\begin{figure}
\epsscale{0.85}
\begin{center}
\plotone{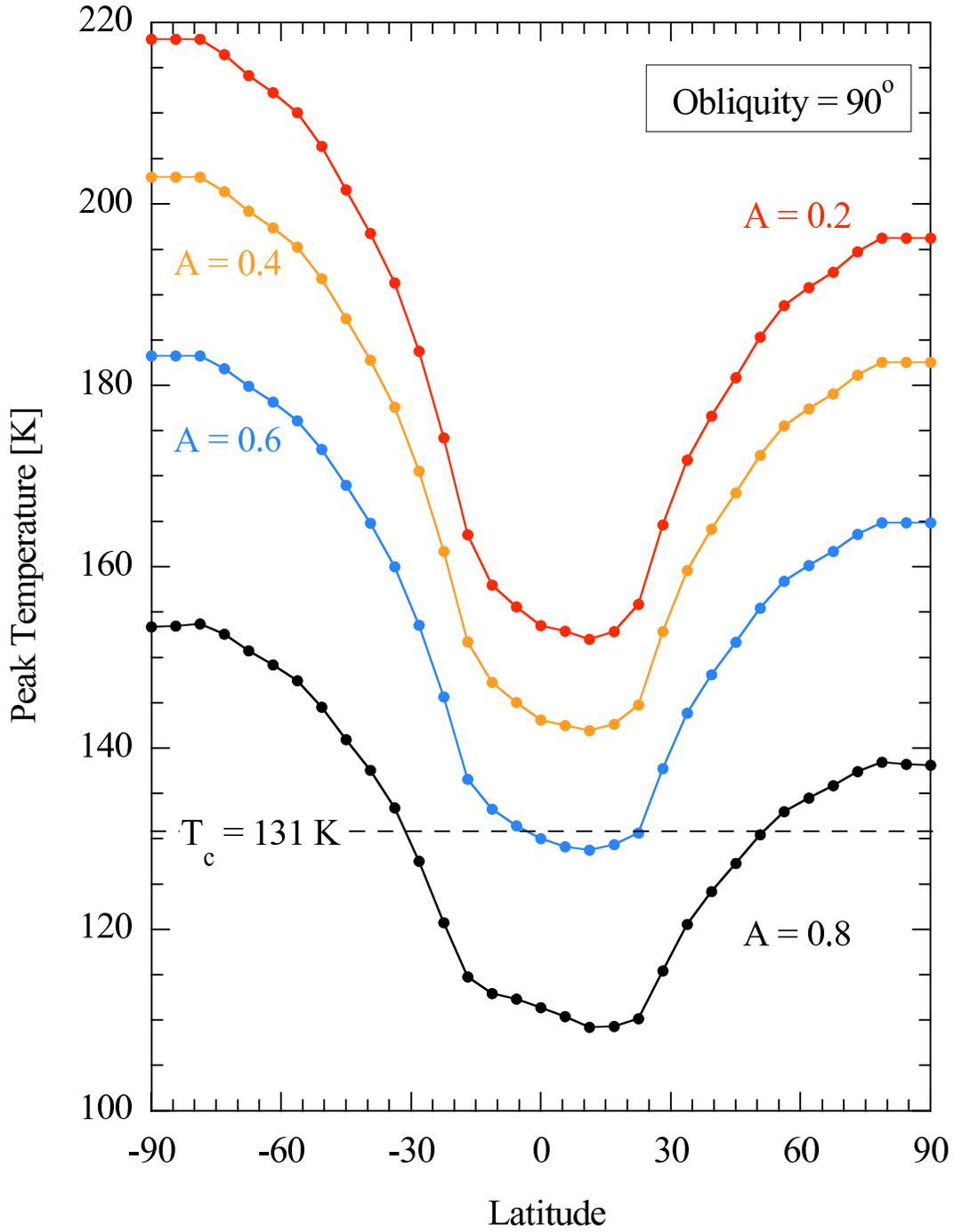}

\caption{Peak temperature as a function of latitude for Bond albedos $A$ = 0.2, 0.4, 0.6, and 0.8 on a spherical model body moving in the orbit of Cybele with obliquity $\psi$ = 90$\degr$.  The critical peak temperature, below which exposed ice can survive for more than 10$^3$ yr, is also marked.    \label{temp_vs_lat} } 
\end{center} 
\end{figure}

To explore the stability of surface ice in more detail, we used a numerical (three-dimensional) heat transport model  (Guilbert-Lepoutre et al.~2011) and considered a rotating body having finite thermal conductivity, a realistic rotation period (6 hr) and three obliquities $\psi =$ 0$\degr$, 45$\degr$ and 90$\degr$.  The surface thermal inertia was taken to be $I$ = 15 MKS units (Mueller and Blommaert 2004), emissivity $\epsilon$ = 0.9 and we assumed that the spin vector at perihelion occupied a plane that also contained the Sun.   Ice Bond albedos from 0.2 $\le A \le$ 0.8 were considered.    We computed the temperature on the surface as a function of position and time.  Sample solutions for the peak annual temperature as a function of latitude are shown in Figure (\ref{temp_vs_lat}).   The model results plotted there refer to a spherical body with 90$\degr$ obliquity moving in the orbit of Cybele and with four values of the ice albedo, as marked.  For this large obliquity, the equator shows the lowest temperatures.  For albedos $>$0.6, the equatorial regions never rise above 131 K, the critical temperature for retaining surface ice over the 10$^3$ yr collision interval.

\begin{figure}
\epsscale{0.85}
\begin{center}
\plotone{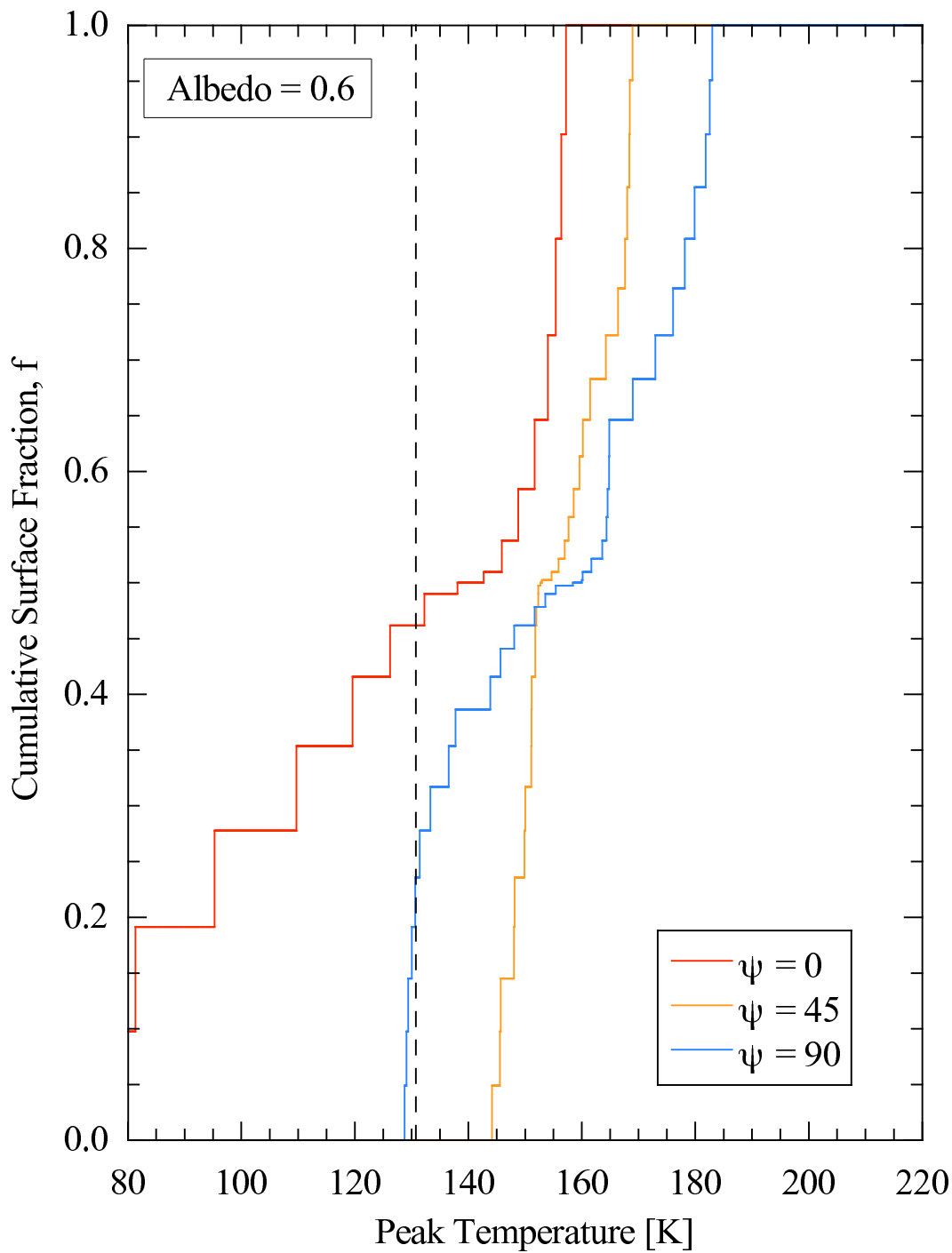}
\caption{Cumulative distribution of the peak temperature calculated over the surface of a rotating sphere following the orbit of Cybele, for three obliquities and ice albedo $A$ = 0.6.  The critical temperature below which exposed water ice can remain for $>$10$^3$ yr is marked by a vertical dashed line.     \label{cuplot_06} } 
\end{center} 
\end{figure}

In Figure (\ref{cuplot_06}) we show the cumulative fraction of the surface whose peak temperature remains less than a given value, $T$.   For simplicity, we show only the results for albedo = 0.6 and tabulate results for other albedos in Table (\ref{conduction}).  A vertical dashed line in Figure (\ref{cuplot_06}) shows the critical temperature $T$ = 131 K (c.f. Figure (\ref{temp_vs_lat})), below which surface ice can survive against sublimation for longer than the 10$^3$ yr collision time.  For the particular case shown, at $\psi$ = 0$\degr$, almost 50\% of the surface remains at $T <$ 131 K all around the orbit, a result of permanently cold  polar regions illuminated obliquely.  Table (\ref{conduction}) shows that, even for $A$ = 0.2, these polar caps survive over a substantial fraction of the surface ($f \sim$ 0.28).   At high obliquity, $\psi$ = 90$\degr$, the cold region shrinks to $\sim$20\% of the surface and is distributed instead as an equatorial band.  The high obliquity models show the highest of all temperatures (at the poles) but retain cold nearly equatorial bands, as shown.  For $A >$ 0.6, these equatorial bands remain below 131 K throughout the orbit and so are potential regions for the preservation of surface ice on 10$^3$ yr timescales.  The smallest temperature variation over the surface is found for  intermediate obliquities ($\psi =$ 45$\degr$).  For $A \le$ 0.6 and this obliquity, no part of the surface is cold enough to retain ice for 10$^3$ yr (Figure \ref{cuplot_06}).

Taken together, the numerical model results show that ice can persist for $\sim$10$^3$ yr timescales on the surfaces of asteroids in Cybele-like and Themis-like orbits, over a wide range of ice albedos and asteroid obliquities.  Small obliquities strongly favor the long-term survival of ice in cold, polar regions, but ice can also survive near the equators of asteroids having obliquities $\sim$90$\degr$.   For albedos $A >$0.6, a millimeter-thick ice layer can survive for $\sim$10$^3$ yr on 10\% or more of the surface regardless of the obliquity.  In this sense, the conduction models are compatible with the scenario proposed above, in which buried ice is excavated by one or more small impacts and spread across a fraction of the asteroid surface as fallback debris.  While we possess no proof that this is the origin of the near infrared spectral features on either Themis or Cybele, the inevitability of impact and the simplicity of the impact excavation model provide ample motivation for an expanded search for ice-like absorptions in, and outgassing from, the asteroids.










%

\clearpage 

\section{Summary}
The near infrared spectral signature of water ice has been recently reported on two main-belt asteroids, (24) Themis and (65) Cybele.  We took optical spectra to search for gaseous emission from these objects caused by the sublimation of surface ice.

\begin{enumerate}
\item The CN 3889\AA~emission line of CN was not detected in either object.  Assuming comet -like H$_2$O/CN mixing ratios, we compute upper limits to the water production rates $<$390 kg s$^{-1}$ for Themis and $<$370 kg s$^{-1}$ for Cybele (both are 5$\sigma$ limits).  These rates are two to three orders of magnitude smaller than outgassing rates expected from the thin, global ice film hypothesis of Rivkin and Emery~(2010).


\item If surface ice exists on Themis and Cybele, it must be clean (albedo $\gtrsim$0.3) and of limited spatial extent (surface fraction  $\lesssim$10\%) in order to simultaneously fit the low average albedos of these objects and the non-detections of gas from sublimated ice. 

\item We speculate that ice might have been exposed on the surfaces of Themis and Cybele by recent impacts. A projectile only 10 m in radius could blanket $\sim$10\% of the surface in ice to a depth of 1 mm, sufficient to produce observable infrared spectral signatures. Impacts onto Themis and Cybele by projectiles of this size  occur on timescales $\sim$10$^3$ yr.


\item Ice on Themis and Cybele with albedo $\gtrsim$0.6 would have a sublimation lifetime in excess of the 10$^3$ yr collision time.  Surface ice, if it is largely free of absorbing contaminants, could be a semi-permanent feature of these and other large, volatile-rich main-belt asteroids.

\end{enumerate}

\acknowledgements
We thank Marc Cassis and Joel Aycock for help with LRIS and Keck and Michal Drahus and the anonymous referee for comments on the manuscript.  The data presented herein were obtained at the W.M. Keck Observatory, which is operated as a scientific partnership among the California Institute of Technology, the University of California and the National Aeronautics and Space Administration. The Observatory was made possible by the generous financial support of the W.M. Keck Foundation.  This work was supported by grants to DJ from NASA Planetary Astronomy and from the NASA Herschel project.

\clearpage

\begin{deluxetable}{llllll}
\tablecaption{Geometrical Circumstances
\label{geometry}}
\tablewidth{0pt}
\tablehead{
\colhead{Name}  & \colhead{UT Date} & \colhead{$R$\tablenotemark{a}} & \colhead{$dR/dt$\tablenotemark{b}}  & \colhead{$\Delta$\tablenotemark{c}} & \colhead{$\alpha$ \tablenotemark{d}}  }
\startdata
24 Themis &  2010 Sep 11 & 3.510 & +0.7 & 2.868 & 14.0  \\
65 Cybele &  2010 Sep 11 & 3.696 & +1.1 & 3.147 & 14.2   \\
\enddata


\tablenotetext{a}{Heliocentric distance [AU]}
\tablenotetext{b}{Rate of change of Heliocentric distance [km s$^{-1}$]}
\tablenotetext{c}{Geocentric distance [AU]}
\tablenotetext{d}{Phase angle [$\degr$]}

\end{deluxetable}

\clearpage

\begin{deluxetable}{llllcr}
\tablecaption{Orbital Properties
\label{orbits}}
\tablewidth{0pt}
\tablehead{
\colhead{Name}   & \colhead{$a$\tablenotemark{a}} & \colhead{$e$  \tablenotemark{b}} & \colhead{$i$\tablenotemark{c}} & \colhead{$q$\tablenotemark{d}}   & \colhead{$Q$\tablenotemark{e}} }
\startdata
24 Themis &   3.129 & 0.134 & 0.76 & 2.708 & 3.548\\
65 Cybele & 3.440 & 0.106 & 3.54 & 3.076 & 3.805 \\
\enddata


\tablenotetext{a}{Semimajor axis [AU]}
\tablenotetext{b}{Orbital eccentricity}
\tablenotetext{c}{Orbital inclination}
\tablenotetext{d}{Perihelion distance [AU]}
\tablenotetext{e}{Aphelion distance [AU]}

\end{deluxetable}

\clearpage

\begin{deluxetable}{llllllcr}
\tablecaption{Physical Properties
\label{physical}}
\tablewidth{0pt}
\tablehead{
\colhead{Name}  & \colhead{$D$\tablenotemark{a}} & \colhead{$p_V$\tablenotemark{b}} & \colhead{$B$\tablenotemark{c}} & \colhead{$B-V$ \tablenotemark{d}} & \colhead{$P$\tablenotemark{e}} & \colhead{$Q_{CN}$\tablenotemark{f}}& \colhead{$dM/dt$\tablenotemark{g}}    }
\startdata
24 Themis &  198$\pm$20\tablenotemark{h} & 0.067$\pm$0.012\tablenotemark{h} & 13.49 & 0.64 & 8.3744$\pm$0.0002 & $\le$3.6$\times$10$^{25}$ & $\le$390 \\
65 Cybele &  273$\pm$12\tablenotemark{i} & 0.050$\pm$0.005\tablenotemark{i}  & 13.64 & 0.68-0.73 & 6.07$\pm$0.05 & $\le$3.5$\times$10$^{25}$& $\le$370  \\
\enddata


\tablenotetext{a}{Diameter in km}

\tablenotetext{b}{Visual geometric albedo}

\tablenotetext{c}{Apparent B-filter magnitude}
\tablenotetext{d}{B-V color}
\tablenotetext{e}{Rotation period in hr}
\tablenotetext{f}{Limit to the CN production rate in number of molecules per second}
\tablenotetext{g}{Limit to the H$_2$O production rate in kg s$^{-1}$}
\tablenotetext{h}{For Themis, we use the value quoted on the JPL Small-Body web site but note that Chernova et al.~(1994) independently reported albedo 0.074}
\tablenotetext{i}{For Cybele, we use measurements by Mueller and Blommaert (2004).  }

\end{deluxetable}


\clearpage

\begin{deluxetable}{lclll}
\tablecaption{Conduction Models\tablenotemark{a}
\label{conduction}}
\tablewidth{0pt}
\tablehead{
\colhead{$\psi$}  & \colhead{$A = 0.2$} & \colhead{$0.4$} & \colhead{$0.6$} & \colhead{$0.8$ }    }
\startdata
0 &  0.28 & 0.35 & 0.46 & 1.00  \\
45&  0 & 0  & 0 & 0.58   \\
90&  0 & 0  & 0.21 & 0.49  \\
\enddata


\tablenotetext{a}{Fraction of the surface having peak temperature $T \le$ 131 K for obliquities 0$\degr$, 45$\degr$ and 90$\degr$ and four values of the ice albedo.}

\end{deluxetable}


\clearpage

\clearpage

\clearpage

\clearpage

\end{document}